\documentclass[12pt]{iopart}

\usepackage{rotating}
\usepackage[colorlinks= true,urlcolor=blue,linkcolor=blue,citecolor=blue]{hyperref} 

\begin{document}

\title[]{Plausible families of compact objects with a Non Local Equation of State}

\author{H. Hern\'andez}
\address{- Laboratorio de F\'isica Te\'orica, Departamento de F\'isica, Universidad de los Andes, M\'erida 5101, Venezuela.}
\address{- Grupo de Investigaciones en Relatividad y Gravitaci\'on, Escuela de F\'isica, Facultad de Ciencias,
Universidad Industrial de Santander, Bucaramanga A.A. 678, Colombia.}
\ead{hector@ula.ve}
\author{L.A. N\'{u}\~{n}ez}
\address{- Grupo de Investigaciones en Relatividad
 y Gravitaci\'on, Escuela de F\'isica, Facultad de Ciencias,
Universidad Industrial de Santander, Bucaramanga A.A. 678, Colombia.}
\address{- Centro Nacional de C\'alculo Cient\'ifico (CeCalCULA), Universidad de Los Andes, 
Corporaci\'on Parque Tecnol\'ogico de M\'erida, M\'erida 5101, Venezuela.}
\ead{lnunez@uis.edu.co}


\begin{abstract}
We investigate the plausibility of some models emerging from an algorithm devised to generate a one-parameter family of interior solutions for the Einstein equations. It is explored how their physical variables change as the family-parameter varies. The models studied correspond to anisotropic spherical matter configurations having a non local equation of state. This particular type of equation of state with no causality problems provides, at a given point, the radial pressure not only as a function of the density but as a functional of the enclosed matter distribution.  We have found that there are several model-independent tendencies as the parameter increases: the equation of state tends to be stiffer and the total mass becomes half of its external radius. Profiting from the concept of cracking of materials in General Relativity, we obtain that those models become more stable as the family parameter increases. 
\end{abstract}
\maketitle
\section{Introduction}

Compact objects are one of the most fascinating entities known in our
Universe. These very particular astrophysical bodies (white dwarfs, neutron stars, quark
stars, hyperon stars, hybrid stars or/and magnetars) seems to be relics of
luminous stars which, in turn, are considered to be born  from extreme events such as
 in supernova explosions. These objects are thought to have cores with densities higher than nuclear
($\rho\approx\left(10-20\right)  \rho_{0}$ with $\rho_{0}\simeq
2 \times10^{14}\ \mbox{gr}\,\mbox{cm}^{-3}$) and enormous gravitational binding energy
($\frac{GM^{2}}{R}\sim5\times10^{53}$ ergs $\sim 0.2 Mc^{2}$). 

The true equation of state that describes the properties of matter at such a high
densities is essentially unknown \cite{Demianski1985}-\cite{Glendenning2000}.
Currently, what is considered to be best understood in this active field, comes from the experimental
insight and extrapolations from the ultra high energy accelerators and cosmic
physics (see \cite{HeiselbergPandharipande2000} and references therein).
Having this uncertainty in mind, it seems reasonable to explore what is
allowed by the laws of physics within the framework of the theory of General Relativity.

We shall consider a general relativistic matter spherical matter configuration, having a particularly 
 \textit{Nonlocal Equation of State}
(\textit{NLES} from now on):
\begin{equation}
P(r)=\rho(r)-\frac{2}{r^{3}}\int_{0}^{r}\bar{r}^{2}\rho(\bar
{r})\ \mathrm{d}\bar{r}
\,\, \Leftrightarrow \,\,
\rho(r) - 3 P(r) + r \left[ \rho(r) - P(r) \right]^{\prime} =0\,,
\label{globalst}
\end{equation} 
which is the static limit of a more general relation between two components of the energy momentum tensor:
\begin{equation}
-\mathbf{T}_{1}^{1}=\mathbf{T}_{0}^{0}-\frac{2}{r}\, \int_0^r \mathbf{T}_{0}^{0}\  \mathrm{d}{\bar{r}}
\,\, \Leftrightarrow \,\,
\left[ {\bf T}_0^0+3{\bf T}_1^1+r\left({\bf T}_0^0+{\bf T}_1^1\right)^{\prime}\right]\left[1-\omega^2\right]=0,
\label{global}
\end{equation}
where $\omega$ is the proper velocity in the radial direction and the prime denote derivative with respect to the radial coordinate $r$ (see  \cite{HernandezNunezPercoco1998, HernandezNunez2004} for details). 

Equation (\ref{global}) was obtained when constant
compactness (i.e. constant gravitational potentials at the surface of the matter configuration) was required for imperfect anisotropic fluid (unequal stresses, i.e. $P\neq P_{\perp}$ and heat flux) matter configurations. It was also found that these type of anisotropic configurations admit a Conformal Killing Vector Field (see reference \cite{HernandezNunezPercoco1998} for details).

It is clear that in equation (\ref{globalst}) a collective behavior on the physical variables $\rho(r)$ and
$P(r)$ is present. The radial pressure $P(r)$ is not only a function
of the energy density, $\rho(r),$ at that point but also its functional
throughout the rest of the configuration. The inner distribution of matters to a given point, 
and not only the density at that point, contributes to the value of the radial pressure. 
It is worth mentioning that there is no possible causality drawback with equation (\ref{globalst})
because it is the static limit of Equation (\ref{global}). When the evolution of a density perturbation is examined, the full dynamical Equation (\ref{global}) has to be considered.

This type of equation of state, originally proposed by D. G. Ravenhall and C. J. Pethick in 1994 \cite{RavenhallPethick1994}, has proven to be very fruitful describing a variety of relativistic astrophysical scenarios \cite{ArrautBaticNowakowski2009, Nicolini2009, HorvatIIijicMarunovic2011A, HorvatIIijicMarunovic2011B}. Particularly, in \cite{ArrautBaticNowakowski2009} the authors, considering the framework of the non commutative geometry, describe a mini black hole having a Schwarzschild geometry outside the event horizon, but whose standard central singularity is replaced by a self-gravitating droplet. The energy-momentum tensor of this droplet is an anisotropic fluid obeying a \textit{NLES}, which allows to have a droplet with a finite radius and positive definite pressure distribution at the interior. More recently, the concept of quasi-local variables was introduced which leads  to a simplified criterion of stability of matter configurations based on the profiles mass $M$ vs radius $R$ \cite{HorvatIIijicMarunovic2011A}.

In General Relativity, static spherically symmetric perfect fluid distributions are described by three independent Einstein equations for four variables (two metric functions, the energy density and the isotropic pressure). As in the case of classical hydrodynamics, in order to integrate the system, additional information has to be provided in the form of an equation of state or an heuristic assumption involving metric and/or physical variables. This situation suggests the possibility to obtain any possible solution by giving a single generating function, inspired in some physical intuitions.  A method to procure solutions in this way has been recently presented by K. Lake \cite{Lake2003} for local spherical isotropic fluid (equal stressed along radial and tangential directions). This algorithm was extended by Herrera and collaborators  \cite{HerreraOspinoDiPrisco2008} in order to consider anisotropic spherical distributions of matter. 

In the literature there are several proposed strategies (with a variety of heuristics) to produce families of solutions having a parameter that generates new family-members (see, for example \cite{StephaniEtal2006}). It is interesting to inquire about how the physical variables change as the parameter is altered and, more specifically, how the variation of the family-parameter affects the stability of the corresponding matter configuration. If the algorithm to generate new solutions leads to plausible models of compact objects and, as the family-parameter varies the family members become better models, thus the proposed method will have more physical meaning.   In this work we shall investigate families of $NLES$ anisotropic matter configuration generated with the Lake-Herrera method by exploring the plausibility of their physical variables and by studying their potential stability when the parameter is changed. 

The structure of the present paper is the following. The next section  sketches the method
to generate families of exact solutions having \textit{NLES}. 
In Section \ref{Examples} we shall workout three examples of families of matter configuration having \textit{NLES}. We close the work with a brief discussion of the results in the Section \ref{conclusion}.

\section{The Method and the \textit{NLES}} 
\label{LakeMethod}

In order to explore the feasibility of \textit{NLES}  in the context of the Lake-Herrera method, we shall use an energy-momentum tensor represented by 
\begin{equation}
\mathbf{T}_{\nu}^{\mu}~=~\mbox{diag}\,~(\rho, -P, -P_{\perp}, -P_{\perp}), \label{MNT}
\end{equation}
to describe static, spherically symmetric, anisotropic (unequal stresses, i.e. $P \neq P_{\perp}$) bounded configurations in General Relativity. 

Following \cite{IshakLake2003}, we can build a set of four independent scalars as: 
$G^\beta_\alpha u^\alpha n_\beta$,  
$G1 \equiv G^\beta_\alpha u^\alpha u_\beta$,  
$G2 \equiv G^\beta_\alpha n^\alpha n_\beta$ and 
$G  \equiv G^\alpha_\alpha$\,.
Where $G^\beta_\alpha$ is the Einstein tensor, with a congruence of unit timelike vectors $u^\alpha = (u^1, u^2, 0, 0)$ and its unit normal field $n^\alpha$ satisfying $u^\alpha u_\alpha = 0$, and $n^\alpha n_\beta = 1$.	

Thus, with the above structure of the energy-momentum tensor, we have the following equations relating those scalars and the physical variables of the configuration:
$
G =8 \pi \left(\rho-P-2P_{\perp} \right)\,,\,\,G1=8\pi \rho \,,\,\,G2 = 8\pi P\,, \,\,-G+G1-G2 =16\pi P_{\perp} \,.
$

For the perfect fluid case, ($P=P_{\perp}$), we can write an equation involving only the above mentioned scalars:
\begin{equation}
-G+G1-3G2=0 \,.
\label{cfp2}
\end{equation}

Considering an static spherically symmetric space time, described by the line element 
\begin{equation}
ds^2 = e^{2 \nu(r)}\,dt^2-e^{2\lambda(r)}\,dr^2-r^2 \left(d\theta ^2+\mbox{sen}^2\theta\,d\phi^2\right)\,.
\label{metricaesferica}
\end{equation} 
If  
$
e^{-2\lambda(r)}=1- 2m(r)/r \, 
$
then, equation (\ref{cfp2}) can be written as 
\begin{equation}
r^2\left(r-2m\right)\left[ \nu'' - (\nu')^2\right]-r\nu'\left[r-3m+rm'\right]-rm'+3m=0 \,. 
\end{equation} 
It is clear that for a given  $\nu (r)$, the equation can be integrated to obtain $ m (r) $, making it possible to build an algorithm to generate families of solutions of Einstein equations in the case of perfect fluid \cite {Lake2003}.
The function $\nu (r)$ is not completely arbitrary because it must fulfill a series of
requirements that prevent singularities on the invariants obtained from the Riemann tensor.

For a perfect, static and spherically symmetric fluid, the fact that the central density
$ \rho_c $ and the central pressure $ P_c $ are finite ensures that all the 
invariants of the Riemann tensor, are regular at the center of symmetry. 

We wish to consider the feasibility of using a
\textit{NLES}  to obtain families of exact solutions of the
Einstein equations through the above mentioned algorithm and to explore the behavior of the physical variables as we vary the parameter that generates the members of the family. 

The metric corresponding to a static spherical  matter configuration having a \textit{NLES} can be written as 
\begin{equation}
ds^2 = e^{2 \lambda(r)}e^{2\kappa}\,dt^2-e^{2\lambda(r)}\,dr^2-r^2 \left(d\theta ^2+
\mbox{sen}^2\theta\,d\phi^2\right)\,,
\label{metriest}
\end{equation}
where  $\lambda(r)=\nu(r) - \kappa$ and  $ \kappa $ is a constant \cite{HernandezNunezPercoco1998, HernandezNunez2004}. 
This metric recalls the so called isothermal coordinates system \cite{Synge1960} which, in turn, is a particular
case of the more general ``warped space-time'' (we refer the reader to
\cite{CarotDaCosta1993} and references cited therein for a general discussion).

With the metric (\ref{metriest}) and the energy momentum tensor (\ref{MNT}), the  
Einstein equations can be written as
\begin{eqnarray}
\rho&=&\frac{1-e^{-2\lambda}\left(1-2 r \lambda^{\prime}\right)}{8\pi r^{2}}\,,\label{enl1est} \\ 
P &=&  \rho - \frac{1-e^{-2\lambda}}{4\pi r^{2}}\,,\label{enl2est} \\
P_{\perp} &=&\frac{\lambda^{\prime \prime}\,e^{-2\lambda}}{8\pi} \,.\label{enl4est}
\end{eqnarray}
Now, combining (\ref {enl1est}) and (\ref {enl2est}) it results that
\begin{equation}
m(r) = 2\pi r^{3}\left(\rho - P  \right) \,.  
\label{eme_est}
\end{equation}

From the matching conditions, at the boundary of the distribution $r=R$, we get: 
\begin{equation}
\kappa=-2\,\lambda_R = \ln(1-2\mu) , \qquad \mathrm{with} \quad \mu =\frac{m(R)}{R}= \frac{M}{R}\,.
 \label{kappa}
\end{equation}
Clearly, from equation (\ref{globalst}) evaluated at the center of the matter distribution, it follows that $\rho_c=3P_c$. 
Additionally, in order to have $P(R)=0$ we must satisfy:
\begin{equation}
2R\lambda^{\prime}_R+1=e^{2\lambda_R}\,, 
\label{conp0}
\end{equation}
where $\lambda_R=\lambda(R)$ and $\lambda^{\prime}_R = \left. \frac{\mathrm{d} \lambda(r) }{\mathrm{d}r} \right|_{r =R}$

\subsection{A family of matter configurations with \textit{NLES} }
Restating a similar algorithm as the one presented in references \cite {Lake2003,HerreraOspinoDiPrisco2008},  we can generate families of exact solutions, for the case of \textit{NLES}, by providing a metric function $\nu(r)$. Our version for the $NLES$ Lake-Herrera method involves the following steps:

\begin{enumerate}
\item Provide a metric function $\nu(r)$ of the form
\begin{equation}
\nu(r)= N \Phi(r) \,, \quad N \geq 1 
\label{nulake22} \,,
\end{equation}
where $ N $ is an integer parameter that defines the family and $ \Phi (r) $ is a function that must be
monotonically increasing with a regular minimum $ r = 0 $ and have to 
fulfill with the matching conditions at the boundary of the distribution \cite{Lake2003}. 

\item From the condition (\ref{conp0})  it follows that 
\begin{equation}
{{\rm e}^{2\,\kappa}}=\left( 1-2\mu \right)^2={\frac {{{\rm e}^{2\,N\Phi _R }}}
{1+2\,RN \Phi'_R }}\,,
\label{acopla}
\end{equation}
with $\Phi_R=\Phi(R)$ and $\Phi^{\prime}_R = \left. \frac{\mathrm{d} \Phi(r) }{\mathrm{d}r} \right|_{r =R}$

\item From the Einstein equations  (\ref{enl1est})-(\ref{enl4est}) we obtain the physical variables: 
\begin{eqnarray}
\rho&=& \frac{1}{8{\pi }{r}^{2}} \, \left[1-\frac {1-2\,rN \Phi'  }{1+2\,RN \Phi_R'}  {\rm e}^{2\,N \left( \Phi_R - \Phi  \right) }\right] \,, \\
P&=& \frac{1}{8{\pi }{r}^{2}} \, \left[-1+\frac {1+2\,rN \Phi'  }{1+2\,RN \Phi_R'}  {\rm e}^{2\,N \left( \Phi_R  - \Phi   \right) }\right]  \,,\\
P_{\perp}&=& \frac{1}{8{\pi }} \, \frac {N \Phi'' }
 { 1+2\,RN \Phi'_R   } {\rm e}^{2\,N \left( \Phi_R -\Phi \right) }\,.
\end{eqnarray}

\item The mass function, equation (\ref{eme_est}), is:
\begin{equation}
m=\frac{r}{2} \left[ 1-{\frac {{{\rm e}^{2\,N \left( \Phi_R -\Phi  \right) }}}
{1+2\,RN \Phi'_R }} \right]\,.
\label{masa_N}
\end{equation}
\item We can also calculate algebraic expressions for the velocities of sound, both radial an tangential as
\begin{equation}
v^2_s= \frac{\partial P}{\partial \rho} =
\frac{ \left[\Psi {r}^{2}-2\,r \Phi' -\frac{1}{N} \right] 
{\rm e}^{2\,N \left( \Phi_R -\Phi \right) } +2\,R\,\Phi'_R  +\frac{1}{N} }
{  \left[  \Psi {r}^{2}+\frac{1}{N}\right] 
 {\rm e}^{2\,N \left( \Phi_R -\Phi  \right) } - \frac{1}{N}-2\,R \, \Phi'_R }\,,
 \label{velr_N}
\end{equation}
\begin{equation}
v^{2}_{s\perp}= \frac{\partial P_{\perp}}{\partial \rho} =
\frac{1}{2} \frac{ r^3\left[  \Phi''' -2\,N  \Phi''  \ \Phi' \right] {\rm e}^{2\,N \left( 2\,\Phi_R -\Phi \right)} }
{  \left[ \Psi {r}^{2}+\frac{1}{N} \right]  {\rm e}^{2\,N \left( 2\,\Phi_R -\Phi \right) } 
-4\,R \left(\Phi'_R +\,{R}N \Phi'^{\,2}_R\right)  -\frac{1}{N} }\,,
 \label{velt_N}
\end{equation}
where $\Psi=\Phi''  -2\,N  \Phi'^{\,2}$.
\end{enumerate}

\subsection{ How this family of \textit{NLES} behaves for Large $N$ ?}
The above method allows us to explore the behavior of the physical variables for different values of  $N$, in particular we can see that for   extreme situations, for large values of $N$, it is clear that:
\begin{equation}
N \gg 1  \quad \Rightarrow 
\rho \sim  \frac {r \Phi' }{R \Phi_R'}  {\rm e}^{2\,N \left( \Phi_R - \Phi  \right) } \quad \mbox{and} \quad
P \sim       \frac {r\Phi'  }{R \Phi_R'}  {\rm e}^{2\,N \left( \Phi_R  - \Phi  \right) }\,,
\label{LargeNPrho}
\end{equation}
thus, the material become very stiff, i.e. when $N \gg 1 $ the equation of state tends to be $\rho=P$, since we know that $ \Phi_R  - \Phi >0$. 

Considering the anisotropy of the material
\begin{equation}
N \gg 1   \quad \Rightarrow 
\frac{P_{\perp} - P}{\rho}
 \sim  \frac { (\Phi'' -2r\Phi') }{2\,r \Phi' }\,. 
 \label{LargeNAnisotropy}
\end{equation}
As it emerges from the hydrostatic equilibrium equation for general relativistic anisotropic fluids, the gradient of pressures is steeper for positive values of the anisotropy ($P_{\perp} - P$). Thus, if $(\Phi''(r) -2r\Phi'(r))>0$ the pressure towards the center will increase more rapidly. It is clear that the effect of the change of the anisotropy as $N$ increases depends on the sign of the equation $(\Phi''(r) -2r\Phi'(r))$ which could be translated into restrictions on the seed function $\Phi(r)$. This will be clear with one of the examples worked out in Section \ref{Examples}.

Now, by evaluating the equation (\ref{masa_N}) at the boundary $r=R$, we obtain for the total mass $M$:
\begin{equation}
N \gg 1   \quad \Rightarrow 
M \sim \frac{R}{2} \,,
\label{LargeNM}
\end{equation}
regardless of the seed function $\Phi(r)$.

Finally, the sound velocities when  $N \gg 1 $ tend to behave like 
\begin{equation}
v^2_s \sim \frac{(\Phi''  -2\,N  \Phi'^{\,2}) {r}^{2}-2\,r \Phi' }{ (\Phi''  -2\,N  \Phi'^{\,2}) {r}^{2}} 
\quad \mathrm{and} \quad
v^{2}_{s\perp}\sim \frac{\Phi''' -2\,N  \Phi''  \ \Phi' }{  (\Phi''  -2\,N  \Phi'^{\,2}) {r}^{2} } \,.
 \label{vellim_N}
\end{equation}
Which are model (seed function) depend. Next section will be devote to present some examples of the above algorithm.

\section{The $NLES$ families by some examples}
\label{Examples}
In this section we shall workout three examples of families of matter configuration that, having \textit{NLES},  emerge from the method above mentioned.
\subsection{ Tolman IV families}
\label{family1}
The Tolman IV
isotropic static solution  was originally presented by R.C. Tolman in
1939 \cite{Tolman1939} and it is also found as a
particular case of a more general family of solution in \cite{Korkina1981} and \cite{Durgapal1982}. It is, in some aspects, similar
to the equation of state for a Fermi gas in cases of intermediate central
densities.

Let us consider the a seed function as:
\begin{equation}
 \Phi(r)=\frac12 \ln\left(a+\frac{r^2}{b} \right) \,,\label{nulake2}
\end{equation}
where $a$ and $b$ are constants.

From (\ref{enl1est}) - (\ref{enl4est}) we get the corresponing physical variables:
\begin{eqnarray}
\rho &=&  \frac{1}{8\pi r^2}\left[\frac{\left(ab+r^2\right)^{(N+1)}-
ab^{(N+1)}e^{2\kappa}+b^Ne^{2\kappa}(2N-1)r^2}{\left(ab+r^2\right)^{(N+1)}}
\right]\,, \quad \label{ee1ejem1} \\
P &=& -\frac{1}{8\pi r^2}\left[\frac{\left(ab+r^2\right)^{(N+1)}-
ab^{(N+1)}e^{2\kappa} - b^Ne^{2\kappa}(2N+1)r^2}{\left(ab+r^2\right)^{(N+1)}}
\right]\,, \quad \label{ee1ejem2} \\
P_{\perp} &=& \frac{Nb^N\left(ab-r^2\right)e^{2\kappa}}{8\pi\left(ab+r^2\right)^{(N+2)}} \,.  \label{ee1ejem3}
\end{eqnarray}

The mass function can also be written as 
\begin{equation}
m(r)=\frac{r}{2}\left[1- \frac{b^N e^{2\kappa}}{\left(ab+r^2\right)^{N}}\right] \,.
\label{emeNtol4}
\end{equation}

Now, the two constants are determined from regularity condition at the origin $r=0$, for the density and pressure, $\rho$ and $P$, i.e. 
\begin{equation}
a= \left(e^{2\kappa}\right)^{\frac{1}{N}}\,,\label{ecparaa}
\end{equation} 
and the constant b is obtained by solving (\ref {acopla}), i.e. 
\begin{equation}
b^N e^{2\kappa}\left[ ab + R^2(2N+1) \right]= \left(ab+R^2\right)^{(N+1)}\,.
 \label{ecparab}
\end{equation}

Another way to see the regularity at the center of these models is to calculate
central density which, in this case can be written as
\begin{equation}
\rho_c=\frac{3Ne^{2\kappa}}{8\pi  b\ e^{\frac{2(N+1)}{N}\kappa}}\, ,
\end{equation}
substituting the corresponding values for $ a $ and $ b $ in the expression for $ m (R) = M $,
Equation (\ref {emeNtol4}), we obtain
\begin{equation}
M=\frac{R}{2}\left[1-\left(\frac{3N}{3N+8\pi R^2 \rho_c} \right)^N \right]\,.
\end{equation}

Below, in Table 3.1, there are several cases for different values of $ N $, where the constants $a$ and $b$ are obtained from Equations (\ref {ecparaa}) and (\ref {ecparab}).
For $ N> $ 4 the calculations will depend on the ability to solve analytically
equation (\ref{ecparab}) but,  in any case, it is always possible to find its numerical solution.

It is important to note  that that $ N = 1 $ corresponds to a solution
Tolman IV-like, which is anisotropic and has a \textit{NLES}. The resulting equations of state for values of $ N> 1 $ have completely different properties, and can appreciated from Figure \ref{fig:FigTolmanIV}.

Table 3.1 shows some numerical values for  compact objects with radius $ R = $ 10 km. 

\begin{center}
\begin{tabular}
[c]{|c|c|c|c|c|c|c|}\hline
\textbf{Model} &$\mu $ & $M (M_{\odot})$&$a$ & $b/M$ & $\rho_{R}$ $\times$ $10^{14}$ (\footnotemark[1]) & $\rho_{c}$ 
$\times10^{15}$(\footnotemark[1])      \\ \hline \hline
$N=1$ &0.25 & 1.69 & 0.25 & 256 & 5.36 & 1.61     \\\hline
$N=2$ &0.29 & 2.00 & 0.41 &122  & 6.32 & 1.81  \\\hline
$N=3$ &0.31 & 2.12 & 0.52 &  97  & 6.72 & 1.87   \\ \hline 
$N=4$ &0.32 & 2.19 & 0.59 &  91  & 6.94 & 1.91   \\ \hline \hline
\end{tabular}\\
{Table 3.1} 
\footnotetext[1]{In $\mbox{gr.cm}^{-3}$}
\end{center}

The equation of state $P=P(\rho)$ for some values 
of $N$ are displayed in Figure \ref{fig:FigTolmanIV}a,  with the original Tolman IV like equation of state for the case $N = 1$. 
We can appreciate a satisfactory behavior of the three profiles for the physical variables  in Figure   \ref{fig:FigTolmanIV}b and also form the anisotropy $(P_{\perp}-P)/\rho$  for different values of the parameter $N$.

\subsection{Families generated from conformally flat solutions}
\label{family2}
Now we shall present two families of models generated from anisotropic conformally flat seed solutions.  The two conformally flat seed solutions considered in this section were presented by B.W. Stewart \cite{Stewart1982}. In order to build the family of nonlocal solution, we shall proceed in two ways. First (Case 1), we will select as a seed function, one inspired by a static conformally flat solution and we turn it to be nonlocal by the use of the method. Secondly (Case 2), we will choose a static solution which is both, conformally flat and nonlocal. In both cases only for $N=1$ the configurations are conformally flat and, as it will become evident, when $N$ increases the behavior of the physical variables for these to models are qualitatively different.

Now, considering  \cite{Stewart1982} we have 
\begin{equation}
\Phi'(r) \equiv r^{-1} - (r^2-2rm(r))^{-1/2} \,, \label{phi_stew}
\end{equation}
as a condition to be satisfy by the metric function
\begin{equation}
e^{2 \nu(r)}= e^{2\Phi}\left(a +  b r^2 e^{-2\Phi}\right)^2 \,,
\end{equation}
and the quantities $ a $ and $ b $ are constants to be determined
by boundary conditions.

Following \cite{Stewart1982}, we start with some known static metric function, $ m (r) $, and by integrating equation (\ref {phi_stew}) 
 the other metric function is obtained. This is the most critical part of this approach and from
the four examples shown in \cite {Stewart1982},  two of them have to do with solutions by series for $ \Phi (r) $.

\subsubsection{Case 1:}
 
Consider Example 1 of \cite {Stewart1982} and take mass function shown there
\begin{equation}
m(r)=\frac{r}{2}\left(1-\frac{{\sin}^2(Kr)}{K^2r^2} \right)
\quad \Rightarrow \quad \lambda=\ln\left(\frac{Kr}{{\sin}(Kr)} \right)\,,\label{lambdaej1}
\end{equation}
where $ K $ is a constant.
From (\ref {phi_stew}) it is obtained
\begin{equation}
e^{\Phi(r)}=\frac{1}{2}Kr\cot\left(\frac{Kr}{2}\right) + C\,, \label{Phi1}
\end{equation}
and if $ e^{2 \Phi(0)} = 1 $ then $ C =  0$.

For simplicity, here we will make the following change of variables
\begin{equation}
f(r)\equiv \cot\left(\frac{Kr}{2}\right)\,, \quad f'=-\frac{K}{2}\left(1+f^2 \right)\,, \quad
f''=\frac{K^2}{2}f\left(1+f^2 \right)\,,
\end{equation}

We want then to study the feasibility of the method taking the following family of metrics
\begin{equation}
\Phi(r)= \ln\left[\frac{K\Pi fr}{2} \right] \,,\quad \mbox{with} \quad
\Pi \equiv a+\frac{4b}{K^2 f^2}  \quad \mbox{and} \quad a=e^{\kappa/N}\,.
\end{equation}

The field equations arising from the nonlocal condition are
\begin{eqnarray*}
\rho&=&\frac{1}{8\pi r^{2}}\left[ 1-e^{-2N\Phi}e^{2\kappa} \left\{1- \frac{N \left[K\Pi f^2 \left(2f- Kr(1+f^2)\right) + 8rb\left(1+f^2\right) \right]  } {K \Pi f^3}
\right\} \right],\\
P &=&  \rho - \frac{1-e^{-2N\Phi}e^{2\kappa} }{4\pi r^{2}} \qquad \mathrm{and} \qquad
P_{\perp} =\frac{N\Phi^{\prime \prime}\,e^{-2N\Phi}e^{2\kappa} }{8\pi},
\end{eqnarray*}
and the mass function
\begin{equation}
m(r)=\frac{r}{2}\left[1-4^N e^{2\kappa}\left[\frac{1}{K\Pi r f}\right]^{2N} \right]\,.
\label{mcaso1}
\end{equation}
From the boundary condition $ P(R) = 0 $ is possible to obtain the constant $ b $:
\begin{equation}
b=\frac{K^2 F^2 e^{\kappa/N}}{4} \left[
\frac{NKR\left(1+F^2\right)-\left(2N+1-e^{-\kappa}\right)F}
{NKR\left(1+F^2\right)+\left(2N+1-e^{-\kappa}\right)F} \right] \,,\label{bp0}
\end{equation}
where $F\equiv \cot\left(\frac{KR}{2}\right)$. 
From $m(R)=M$ it is obtained 
\begin{equation}
4Rb=KF\left[2-KRF e^{\kappa/2N} \right] e^{\kappa/2N}\,. \label{bemeM}
\end{equation}
From equations (\ref{bp0}) and (\ref{bemeM}) there results the following transcendental equation for the factor $ KR $
\begin{equation}
2KRFe^{\frac{\kappa}{2N}}= 
\pm \sqrt {\frac{ 
{F^2\left(4\left[2N+1-e^{-\kappa} \right] e^{\frac{\kappa}{2N}}+N\right)  + N  }}{N\left[1+F^2\right] } } +1 \,.
\end{equation}
For a fixed value of $KR$ and different values of $N$,  it is possible to obtain a numerical 
solution of this equation for the amount  $\mu$. The constant $b$ is then obtained for (\ref{bp0}).

Following, the central density can be calculated for all $ N $
\begin{equation}
\rho_c = \frac{N}{16 \pi} \left[12be^{-\kappa/N}-K^2\right] \,.
\end{equation}
By combining this equation with (\ref{mcaso1}) evaluated at the surface is:
\begin{equation}
M=\frac{R}{2}\left[1-\left(\frac{6NKF}{R\left[NK^2\left(3F^2+1\right)+16\pi \rho_c \right]} \right)^{2N} \right]\,.
\end{equation}

The following table shows some numerical values for
a compact object of radius $ R=10$ km and $KR\ $= 0.1
\begin{center}
\begin{tabular}
[c]{|c|c|c|c|c|c|c|}\hline
\textbf{Model} &$\mu$ &$M(M_{\odot})$&$a$ & $bM^2$ & $\rho_{R}$ $\times$ $10^{14}$(\footnotemark[1]) & $\rho_{c}$ $\,\times10^{15}$(\footnotemark[1])   \\ \hline \hline
$N=1$ &0.30 & 2.00& 0.41 &  0.02 & 6.33 & 1.81   \\\hline
$N=2$ &0.32 & 2.19& 0.59 &  0.02 & 6.94 & 1.91 \\\hline
$N=3$ &0.33 & 2.27& 0.69 &  0.02 & 7.17 & 1.95  \\ \hline 
$N=4$ &0.34 & 2.30& 0.75 &  0.01 & 7.29 & 1.97 \\ \hline \hline
\end{tabular}\\
{Table 3.2} 
\footnotetext[1]{In $\mbox{gr.cm}^{-3}$}
\end{center}

Figure \ref{fig:FigCase1} shows a set of graphs for models here referred to as the Case 1 of 
conformally flat nonlocal configuration. 
The equation of state $P=P(\rho)$ for some values 
of $N$ can be appreciated in Figure  \ref{fig:FigCase1}a,  as well as the equation of state 
$P=\rho$ when N = 1 which is the original member of the family.  We can also appreciate   $ (P_{\perp}-P)/\rho$  for each $N$ in Figure \ref{fig:FigCase1}b.

\subsubsection{Case 2:}

A completely different treatment of the above is to start from
\begin{equation}
e^{2\lambda}= K^2 r^2 \left[ 1 + \mbox{cot}^2 \left( Kr+A\right)  \right] =
\left[\frac{K r}{\sin\left(Kr+A\right)}\right]^2  \,,\label{weyl1ocal}  
\end{equation}
where $ K $ and $ A $ are integration constants.

The constant A should vanish because it is necessary that the metrical elements are regular functions at the origin, i.e.
\begin{equation}
A=0 \quad \Rightarrow \quad \lim_{r \rightarrow 0} e^{2\lambda}=1 \,.
\end{equation}
It is clear that, by imposing another condition on the metric elements, it is possible to  completely determine the system of equations.

Now reapplying the algorithm we get
\begin{equation}
\Phi(r)=\ln\left[\frac{K r e^{\kappa} }{{\sin}(Kr)} \right] \,.
\end{equation}
Again, in order to $e^{2\lambda(0)}=1$, we must have $
\lambda(r)=N \Phi-N\kappa\,.
$

From the field equations we get
\begin{eqnarray}
\rho&=&\frac{1}{8\pi r^2} \left[ \left[2N-1 -\frac{NKr\ {\sin}(2Kr)}{{\sin}^2(Kr)}\right] \left[\frac{{\sin}(Kr)}{K r } \right]^{2N} + 1 \right],  \label{enlcp1} \\
P&=&\frac{1}{8\pi r^2} \left[ \left[2N+1 -\frac{NKr\ {\sin}(2Kr)}{{\sin}^2(Kr)}\right] \left[\frac{{\sin}(Kr)}{K r } \right]^{2N} - 1 \right], \label{enlcp2}\\
P_{\perp} &=& \frac{N}{8\pi r^2} \left[\frac{K^2r^2}{{\sin}^2(Kr)} -1\right]\left[\frac{{\sin}(Kr)}{K r } \right]^{2N}\,, \label{enlcp3}
\end{eqnarray}
and the corresponding mass function 
\begin{equation}
m(r)=\frac{r}{2}\left[1- \left(\frac{{\sin}(Kr)}{K r } \right)^{2N} \right].
\label{mcaso2}
\end{equation}
At the surface where $ m (R) = M $, we should satisfy the following identity
\begin{equation}
\left(\frac{{\sin}(KR)}{K R } \right)^{2N}=e^{\kappa} \quad \Rightarrow \quad
{\sin}(KR) = KR e^{\left(\frac{\kappa}{2N}\right)}\,,
\label{coneme}
\end{equation}
and the other condition at the surface, $ P (R) = 0 $, is
\begin{equation}
\left[2N+1 -\frac{NKR\ {\sin}(2KR)}{{\sin}^2(KR)}\right] \left[\frac{{\sin}(KR)}{K R } \right]^{2N} - 1 =0\,.
\label{condpe}
\end{equation}
But by using (\ref{coneme}) and (\ref{condpe}) it is obtained
\begin{equation}
K^2 R^2 =  e^{\left(-\frac{\kappa}{N}\right)}  - 
\left(\frac{2N+1-e^{-\kappa}}{2N}\right)^2\,.\label{elkw0}
\end{equation}
This expression must be substituted into the transcendental equation (\ref {coneme}) in order to obtain (numerically)  a value of the mass-radius ratio $ \mu $ for different values of $ N $.

Then, the central density turns out to be
\begin{equation}
\rho_c= \frac{NK^2}{8\pi} \,.
\end{equation}
Combining this last equation with (\ref{mcaso2}), evaluated at the surface, it
results in
\begin{equation}
M=\frac{R}{2}\left[1-\left(\frac{N}{2\pi R^2 \rho_c} \right)^N \left({\sin}\left[\sqrt{\frac{2\pi \rho_c}{N}}R\right] \cos\left[\sqrt{\frac{2\pi \rho_c}{N}}R\right] \right)^{2N}\right]\,.
\end{equation}

The following table shows some numerical values for a compact object of radius $ R = 10 $  km.

\begin{figure}[ht]
\begin{sideways}
{\includegraphics[height=8.5cm, width=9.4cm]{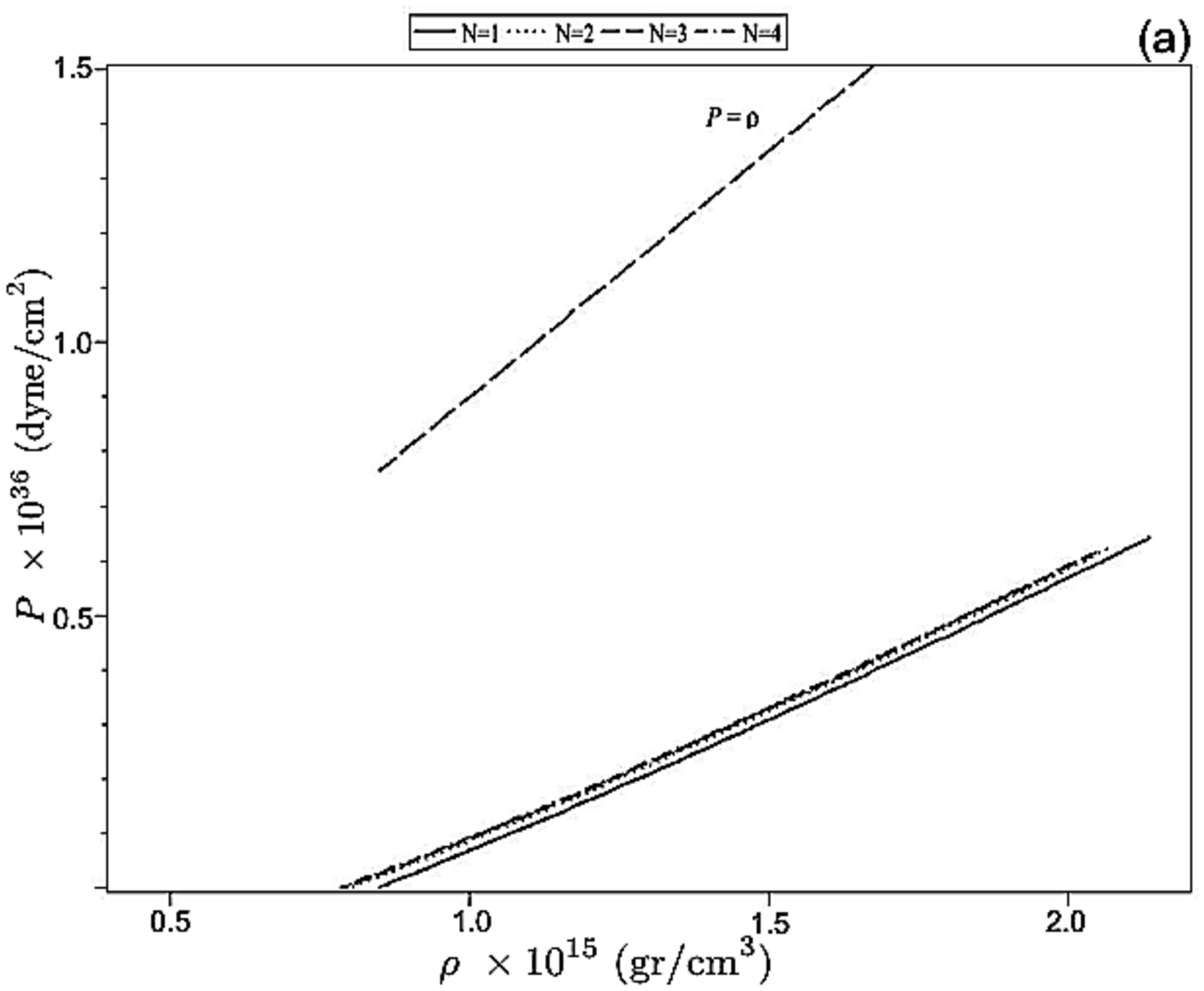}} \hspace{0.1 cm}
{\includegraphics[height=8.5cm, width=9.4cm]{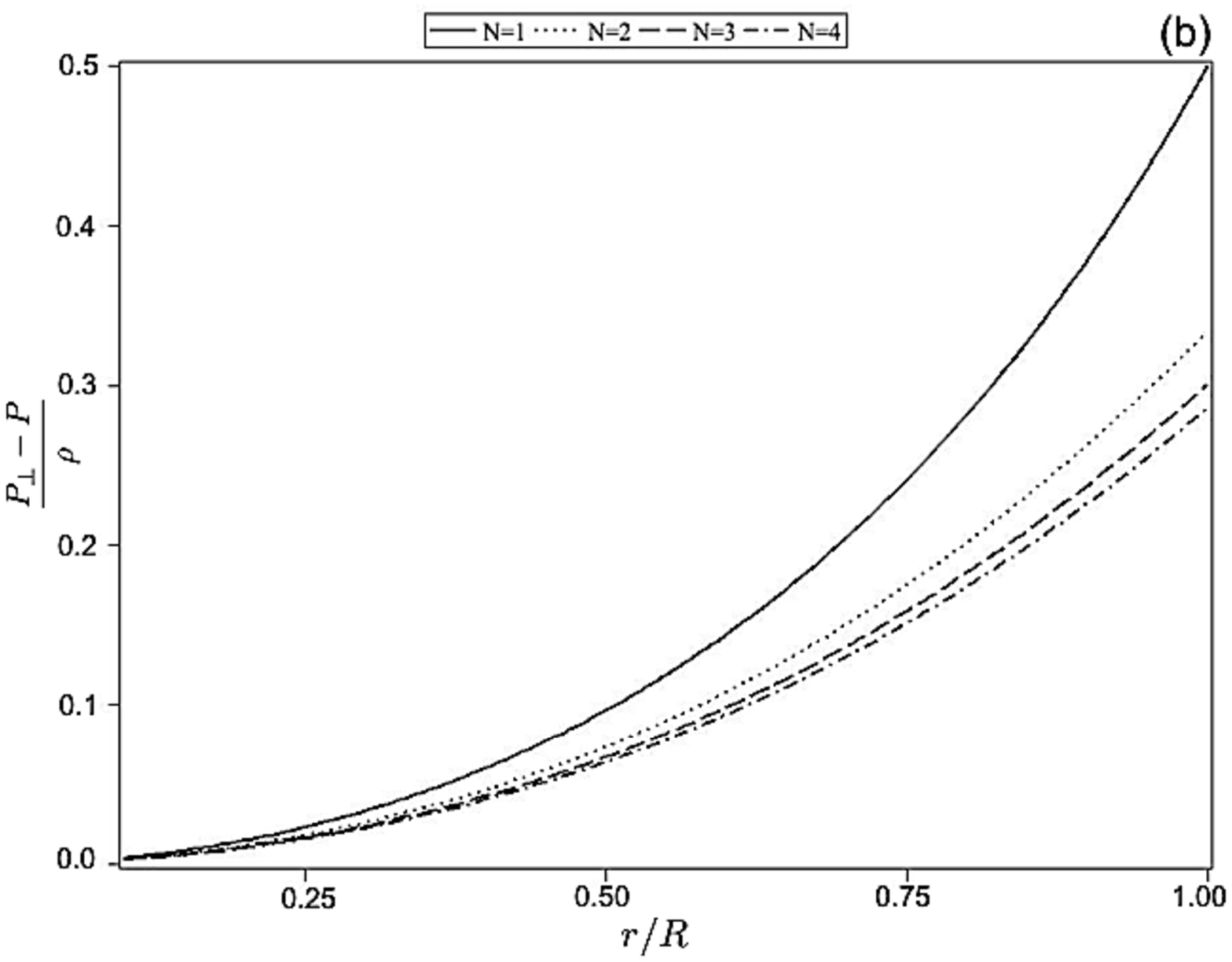}} \\ 
\end{sideways}
\begin{sideways}
{\includegraphics[height=8.5cm, width=9.4cm]{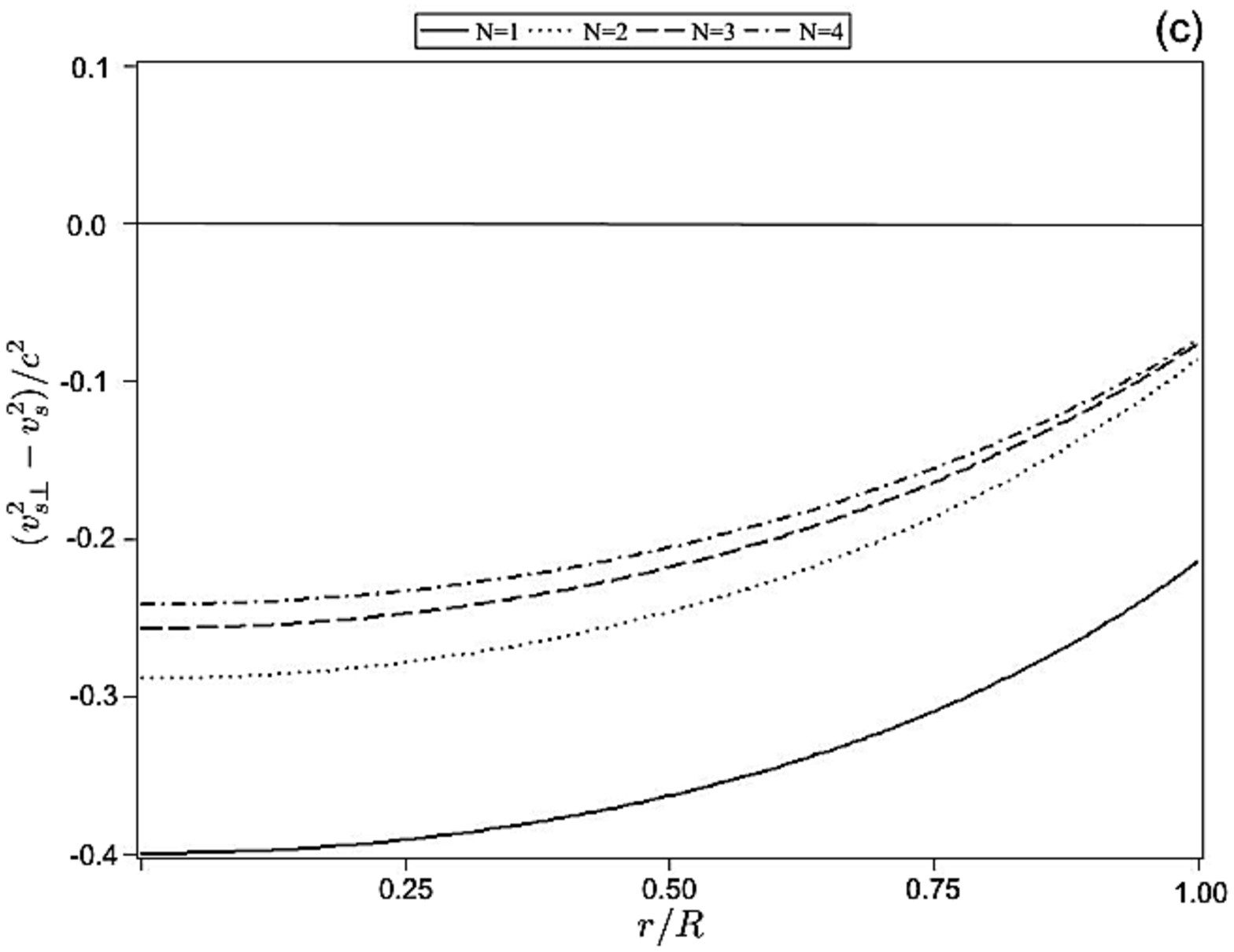}} \hspace{0.1 cm}
{\includegraphics[height=8.56cm, width=9.4cm]{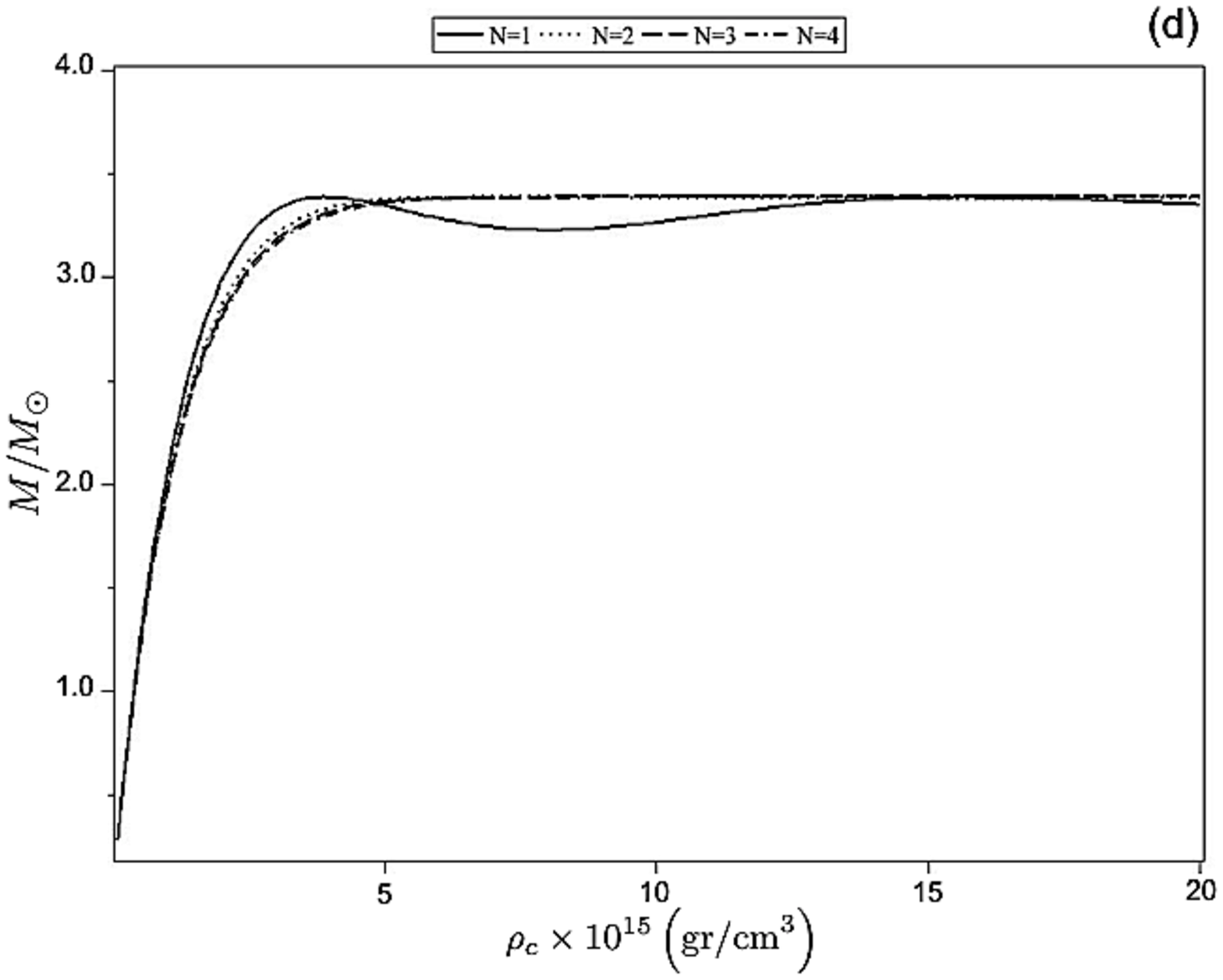}} 
\end{sideways}
\caption{Case 2}
\label{fig:FigCase2}
\end{figure}

\begin{center}
\label{table3}
\begin{tabular}
[c]{|c|c|c|c|c|c|}\hline
\textbf{Model} & $\mu$ & $M_{\odot}$ & $KM$ &$\rho_{R} \times 10^{14}$(\footnotemark[1]) & $\rho_{c} \times10^{15}$(\footnotemark[1])    \\ \hline \hline
$N=1$ & 0.396 & 2.683& 0.79 & 8.49 & 2.14   \\\hline
$N=2$ & 0.374 & 2.535 & 0.52 & 8.03 & 2.07  \\\hline
$N=3$ & 0.368 & 2.494& 0.42 & 7.90 & 2.05  \\ \hline 
$N=4$ & 0.366 & 2.475& 0.36 & 7.84 & 2.04  \\ \hline \hline
\end{tabular}\\
{Table 3.3} 
\footnotetext[1]{In $\mbox{gr.cm}^{-3}$}
\end{center}
It is clear that for values of $ N \neq 1 $ 
Figure \ref{fig:FigCase2} shows a set of graphs for models here referred to as Case 2. 
The equation of state $P=P(\rho)$ for some values 
of $N$ are sketched in Figure \ref{fig:FigCase2}a,  as well as the equation of state $P=\rho$. 
We can appreciate  the anisotropy  $ (P_{\perp}-P)/\rho$  for different $N$ in Figure \ref{fig:FigCase2}b.

\subsection{The stability of the models}
Fluctuations in density and anisotropy induce total radial forces which, depending on 
their spatial distribution, may lead to the cracking, i.e. radial force directed inwards, or, overturning, directed 
outwards, of the source \cite{DiPriscoHerreraVarela1997}.  We can  evaluate potentially unstable regions within anisotropic models based on the difference of the propagation of sound within the matter configuration. Those regions where 
the radial sound speed  is less than the  tangential sound speed:
${v}^{2}_{s} <  {v}^{2}_{s\perp}$ could be potentially unstable, but if  ${v}^{2}_{s} >  {v}^{2}_{s\perp}$  everywhere within a matter distribution, no cracking will occur and the configuration will be potentially stable \cite{AbreuHernandezNunez2007b}. 

By means of the equations (\ref{velr_N}) and (\ref{velt_N}) is possible to determine the sign of 
$\delta v_{s}^2\equiv v_{s\perp}^2 - v_{s}^2$ and to explore how it changes as $N$ varies. Figures \ref{fig:FigTolmanIV}c, \ref{fig:FigCase1}c, and \ref{fig:FigCase2}c show different profiles of $\delta v_{s}^2$
for  Tolman IV like, Case 1 and Case 2 models, respectively. 

Notice that the original seed functions $(N=1)$ for models  Tolman IV like and Case 1 are potentially unstable, this 
means that there are regions where cracking could occur, however, as $N$ increases the possibility of 
occurrence of cracking vanishes. Unlike this situation the Case 2  is potentially stable for all $N$.

\section{Conclusions}
\label{conclusion}
In this work we have found three new families of solutions describing nonlocal anisotropic compact objects by applying  the Lake-Herrera  \cite{Lake2003, HerreraOspinoDiPrisco2008} algorithm. As it was illustrated with the physical quantities shown in the above tables 3.1, 3.2 and 3.3, 
\[
0.25 < \mu <  0.39  \qquad 1.60M_{\odot} < M < 2.68M_{\odot}
\qquad 1.61 < \rho_{c} \times10^{15} <  2.04
\]
the models worked out describe plausible compacts relativistic objects. 

\begin{figure}[ht]
\begin{sideways}
{\includegraphics[height=8.5cm, width=9.4cm]{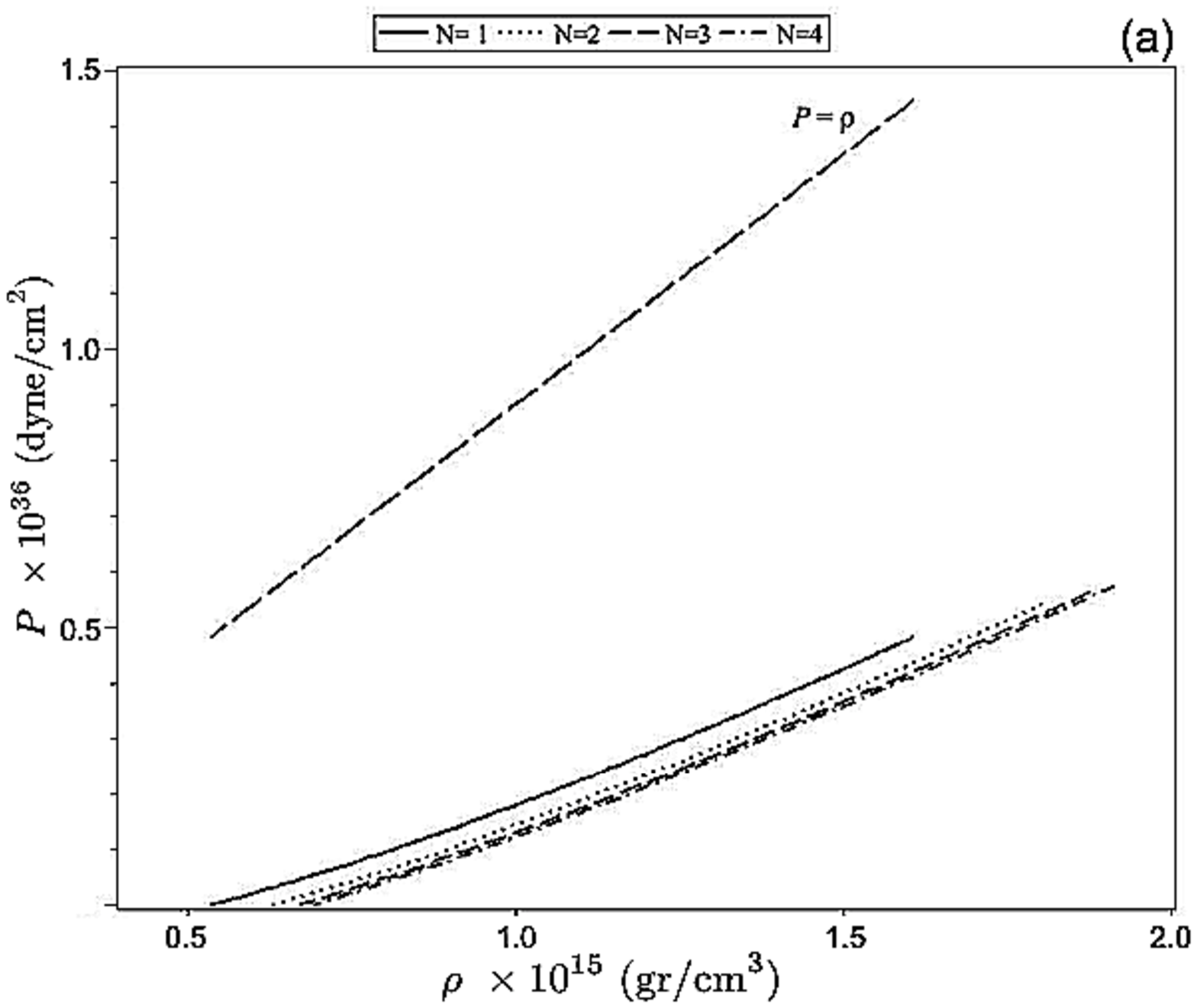}} \hspace{0.1 cm}
{\includegraphics[height=8.5cm, width=9.4cm]{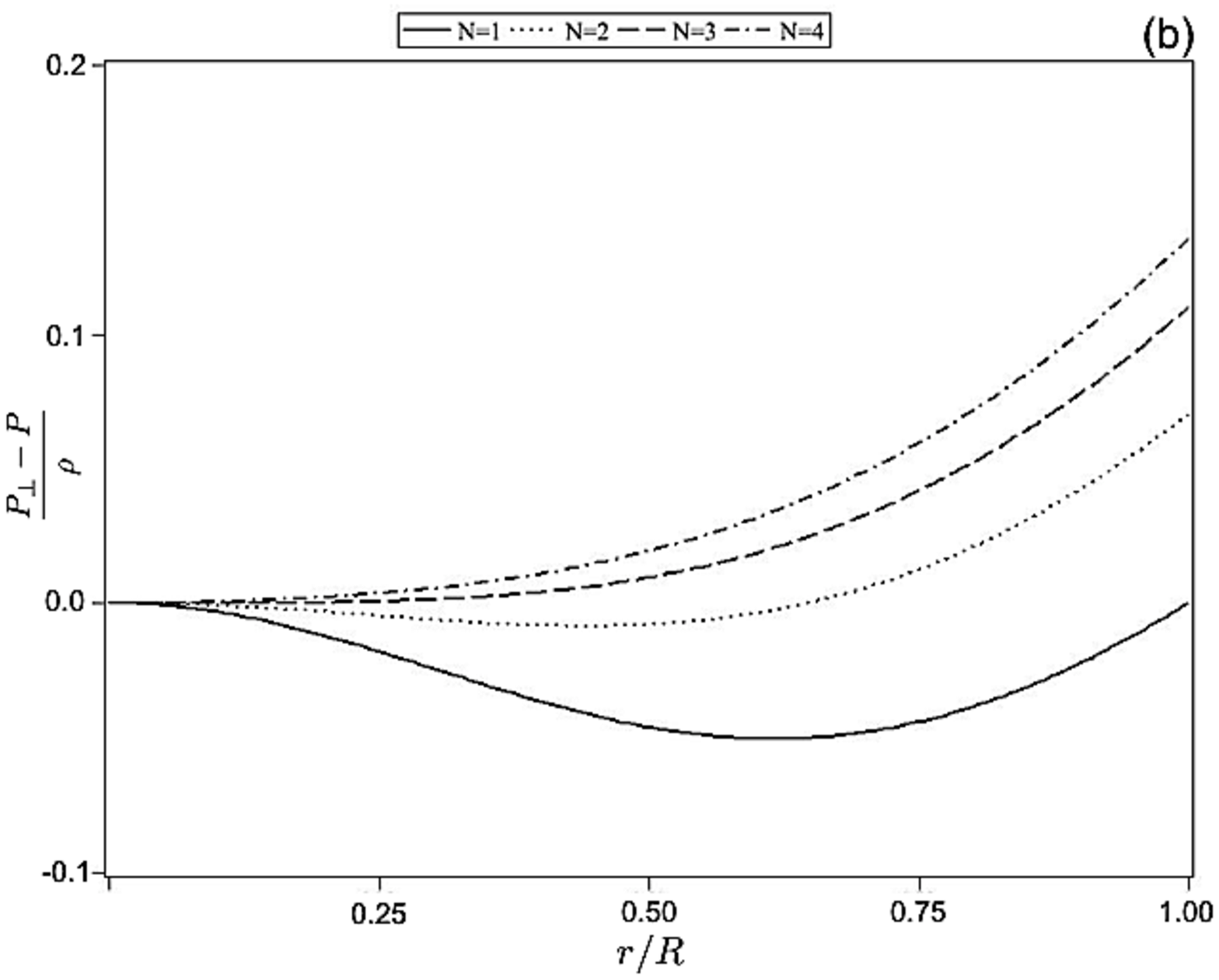}} 
\end{sideways}
\begin{sideways}
{\includegraphics[height=8.5cm, width=9.4cm]{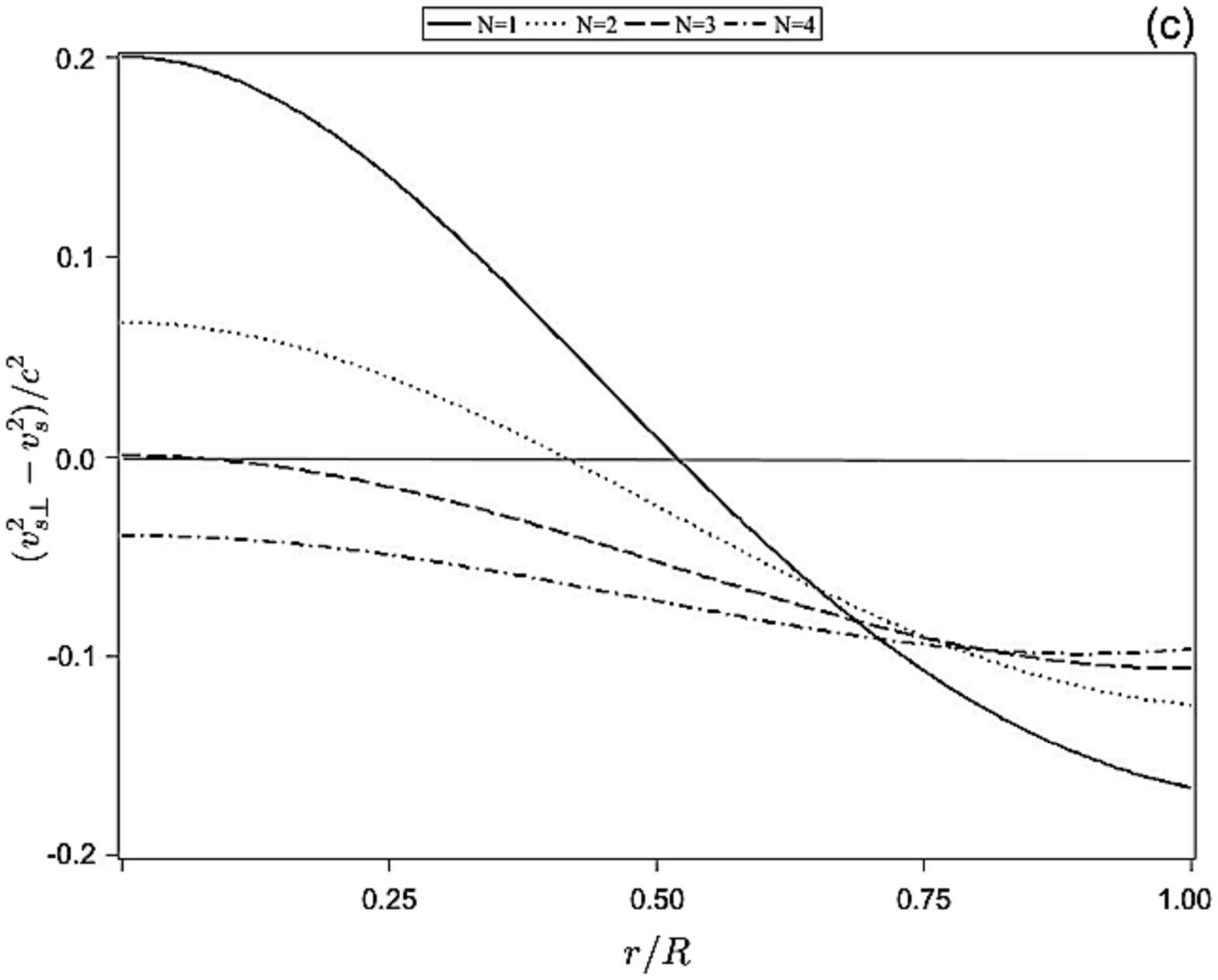}} \hspace{0.1 cm}
{\includegraphics[height=8.5cm, width=9.4cm]{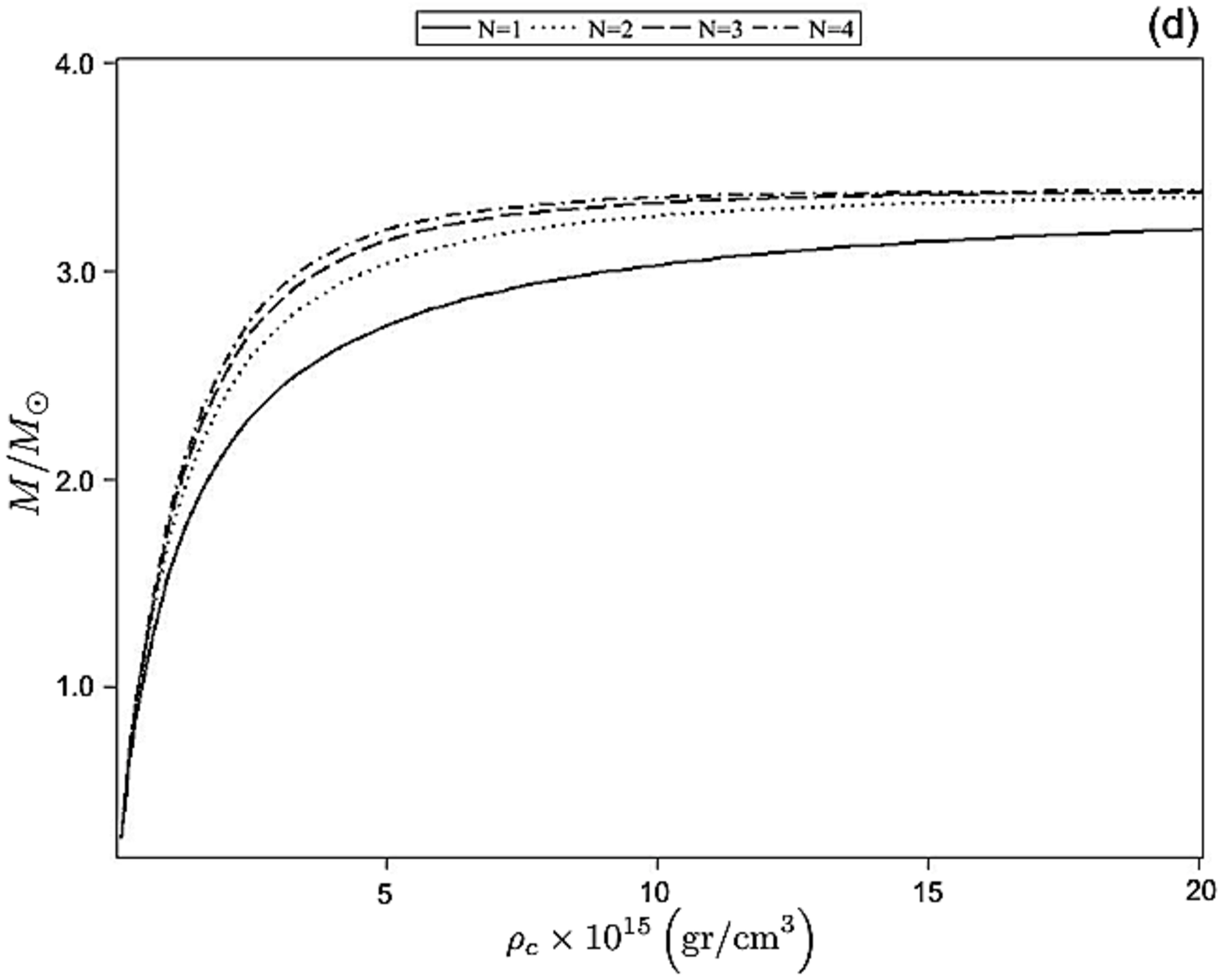}} 
\end{sideways}
\caption{Tolman IV like}
\label{fig:FigTolmanIV}
\end{figure}

\begin{figure}[ht]
\begin{sideways}
{\includegraphics[height=8.5cm, width=9.4cm]{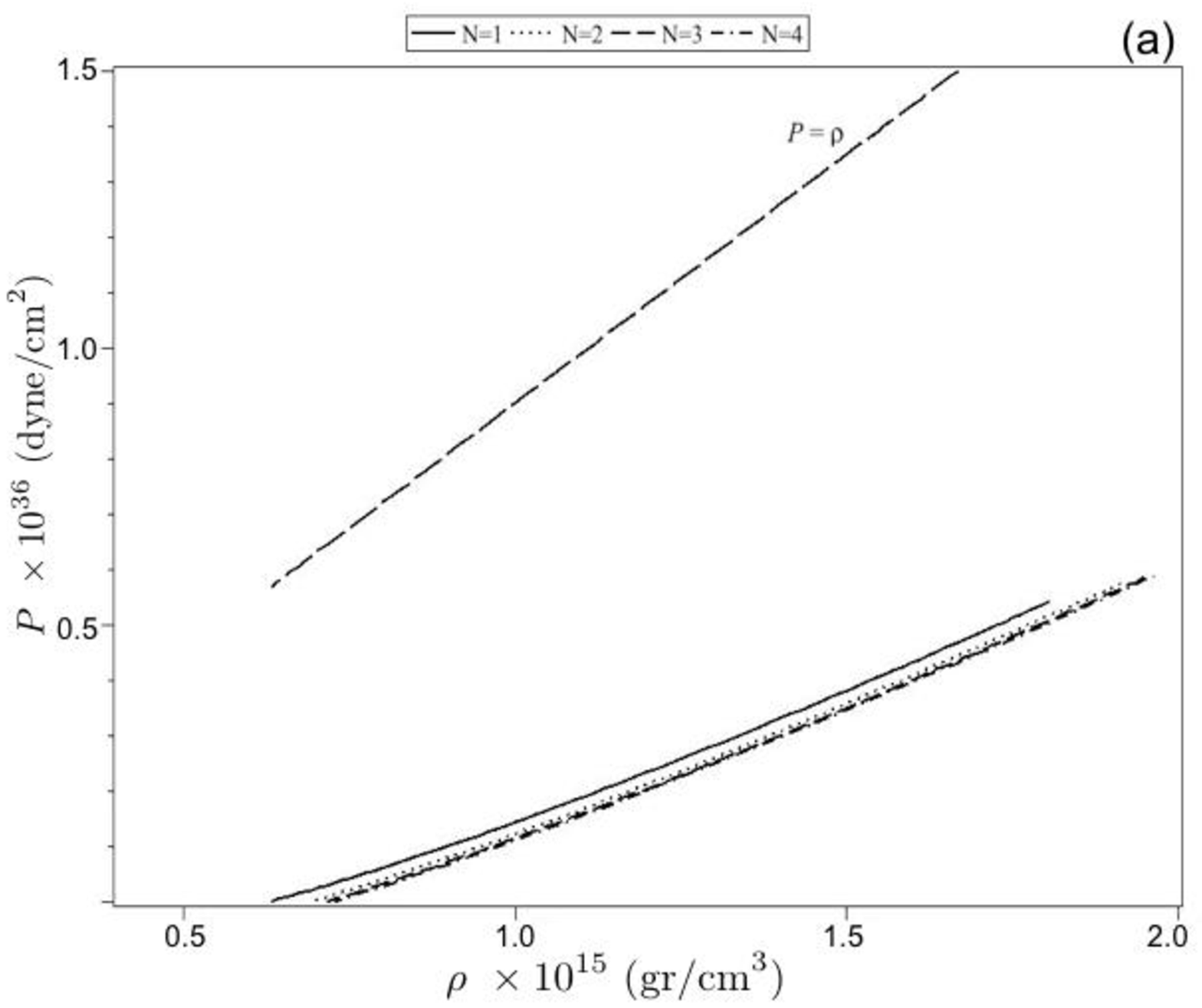}} \hspace{0.1 cm}
{\includegraphics[height=8.5cm, width=9.4cm]{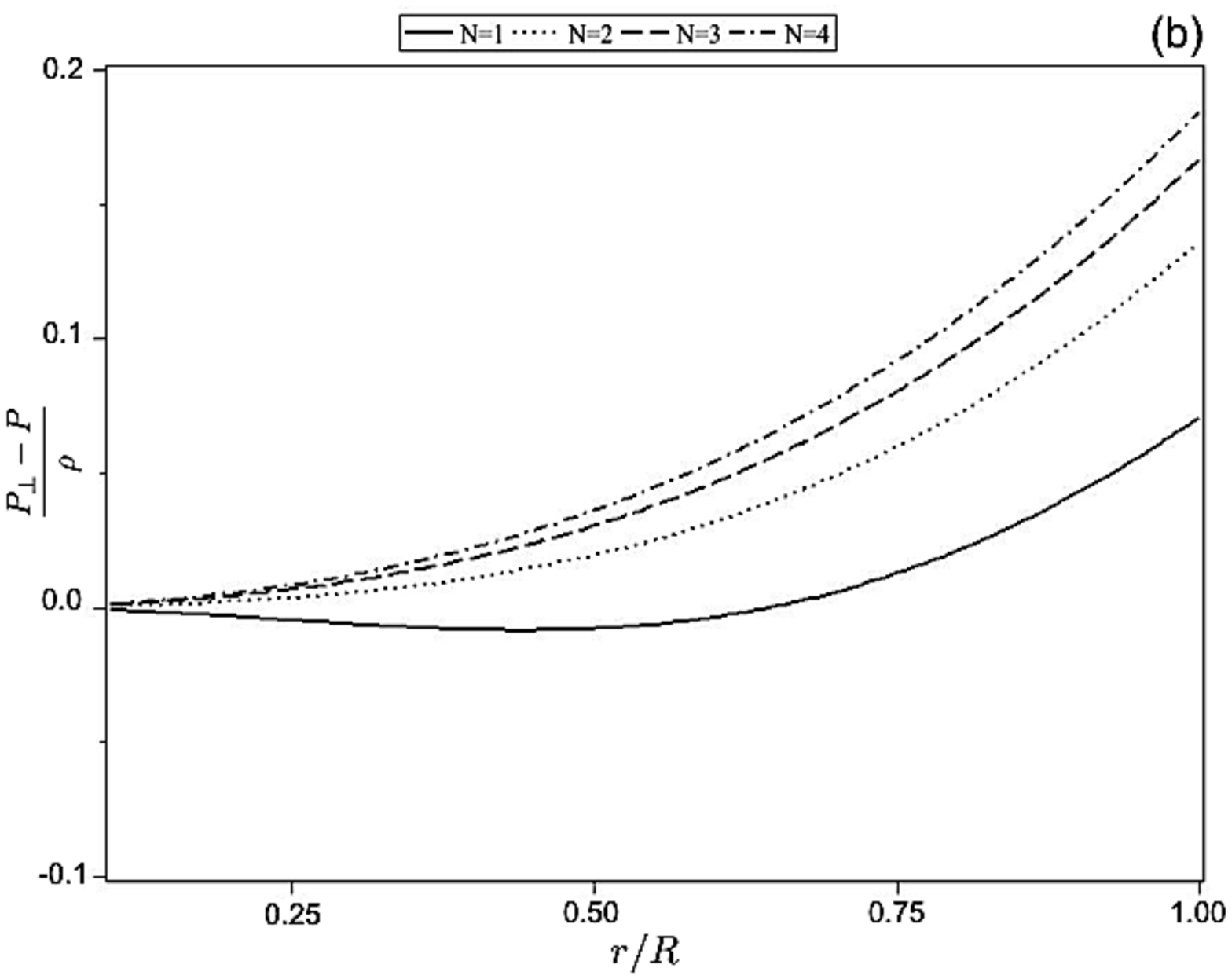}} \\ 
\end{sideways}
\begin{sideways}
{\includegraphics[height=8.5cm, width=9.4cm]{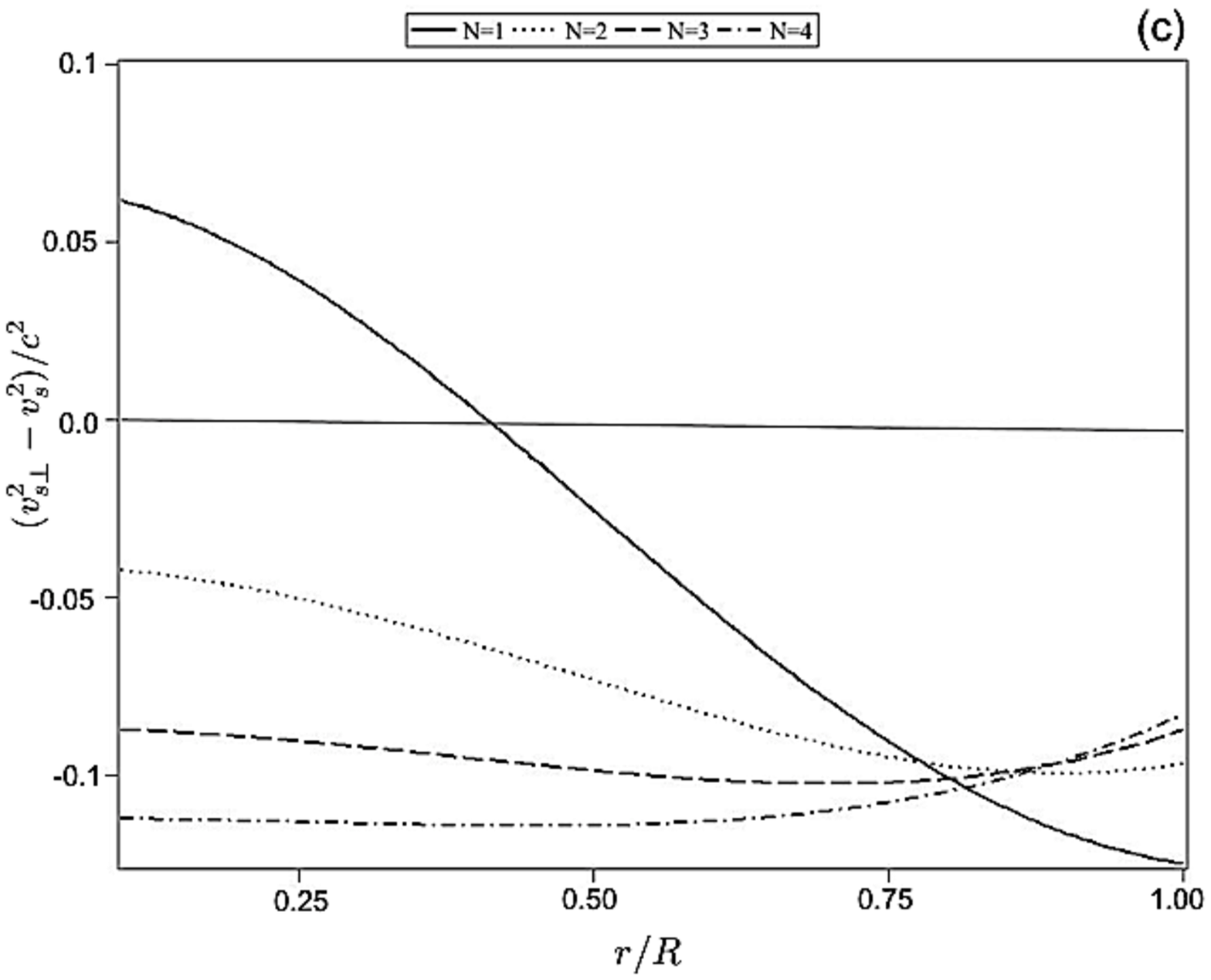}} \hspace{0.1 cm}
{\includegraphics[height=8.5cm, width=9.4cm]{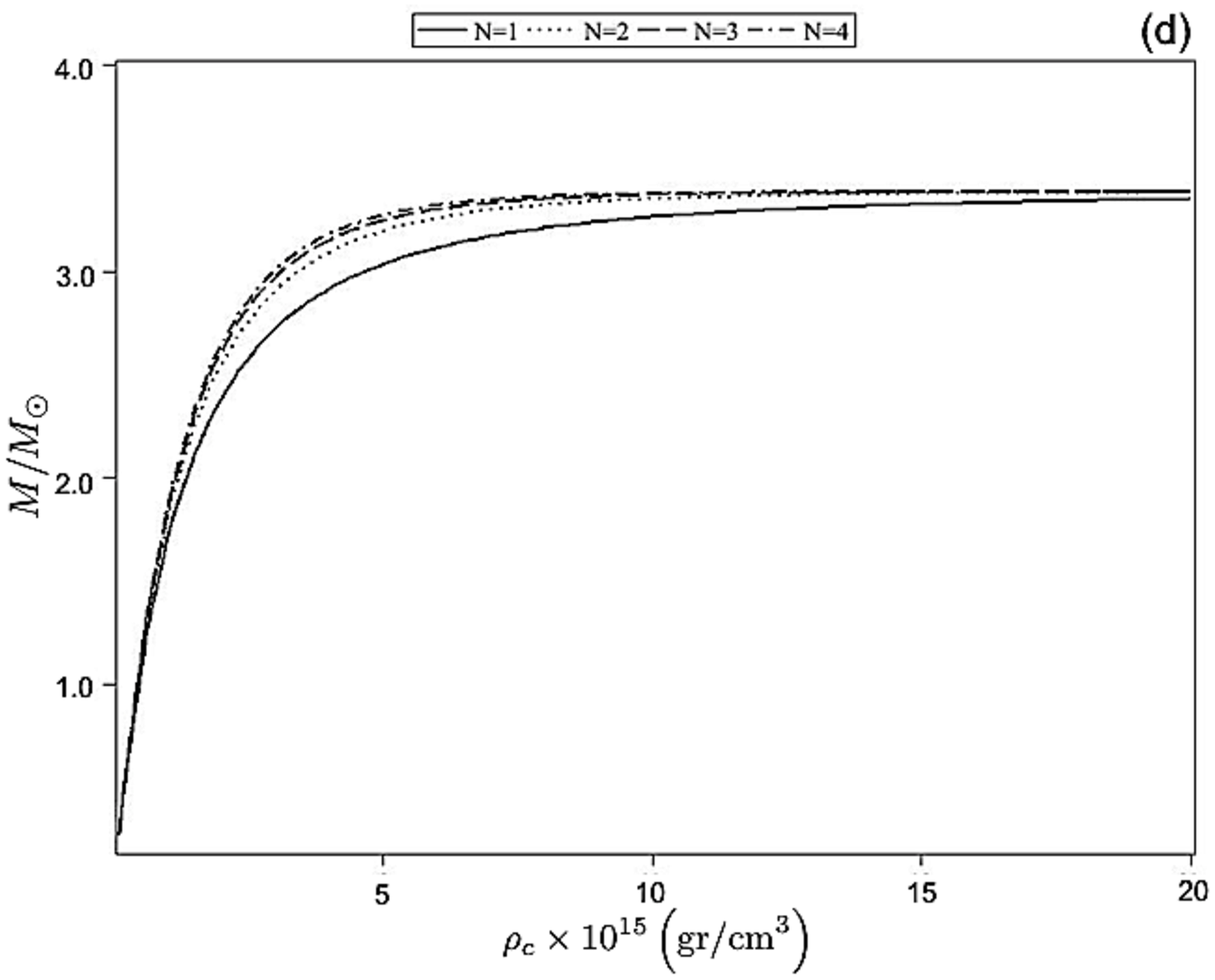}} 
\end{sideways}
\caption{Case 1}
\label{fig:FigCase1}
\end{figure}

With the particular  ansatz (\ref{nulake22}) considered, we have found that:
\begin{itemize}
  \item if $N$ increases the material become stiffer and the equation of state approaches the limit $\rho=P$. This is clear from (\ref{LargeNPrho})  and also it is apparent from the Figures  \ref{fig:FigTolmanIV}a, \ref{fig:FigCase1}a and \ref{fig:FigCase2}a.
  \item the effect of the change in the anisotropy is model dependent.  It depends on the sign of the factor $(\Phi''(r) -2r\Phi'(r))$ which could be translated into restrictions on the seed function $\Phi(r)$. This model dependent behavior could be appreciated comparing figures \ref{fig:FigCase1}b and \ref{fig:FigCase2}b. In the first case, as $N$ increases the anisotropy $ (P_{\perp}-P)/\rho $, for a particular point within the matter configuration, increases but, for the Case 2, as $N$ increases the anisotropy decreases.   
  \item concerning the total mass, $M=m(R)$, it is clear that for $ N \gg 1 $ we got  $ M \sim \frac{R}{2} $ regardless of form of the seed function $\Phi(r)$. This is consistent with the graphs sketched in figures  \ref{fig:FigTolmanIV}d, \ref{fig:FigCase1}d and \ref{fig:FigCase2}d for the three equations of state considered. It is found that total masses approaches a limit lower than  $3.4$  solar masses,  i.e. 
$R/2$ $ \forall$ $N$.
  \item if the initial model of the family presents potential instability, it could become potentially stable as $N$ increases. This is clear from figures  \ref{fig:FigTolmanIV}c and \ref{fig:FigCase1}c where the relation $\delta v_{s}^2= v_{s\perp}^2 - v_{s}^2$ for different values of $N$ are displayed. As it can be appreciated form Figure  \ref{fig:FigTolmanIV}c  $ \delta v_{s}^2 < 0$ for $ N>2 $ and from Figure  
  \ref{fig:FigCase1}c  $ \delta v_{s}^2 < 0$ for $ N>1 $. 
\end{itemize}

For the considered ansatz (\ref{nulake22}), the Lake-Herrera method seams to be a very reasonable algorithm to generate models for compact object with a $NLES$. 

It is worth mentioning that the set of nonlocal solutions obtained with this algorithm describes physically reasonable compact objects. There is no causality drawback for any of the $NLES$ considered, and in fact there is no causality problems for any $NLES$. 
 The Equation (\ref{globalst}) is the static limit of Equation (\ref{global}) and when the evolution of a density perturbation is examined, the full dynamical Equation (\ref{global}) has to be taken into account.

\section*{Acknowledgments}

We gratefully acknowledge the
financial support of the Vicerrector\'ia de Investigaci\'on y Extensi\'on de la Universidad Industrial de 
Santander under project 5541.

\end{document}